\documentclass[aps,prl,twocolumn,showpacs,groupedaddress]{revtex4}

\usepackage{graphicx}
\usepackage{dcolumn}
\usepackage{bm}
\begin{document}


\title{ 
Polyhedral vesicles
}

\author{Hiroshi Noguchi}
\email[]{noguchi@ims.ac.jp}
\affiliation{Department of Theoretical Studies,
Institute for Molecular Science
Okazaki 444-8585, Japan}

             
\begin{abstract}
Polyhedral vesicles with a large bending modulus
of the membrane such as the gel phase lipid membrane were studied
 using a Brownian dynamics simulation.
The vesicles exhibit various polyhedral morphologies such as 
tetrahedron and cube shapes.
We clarified two types of line defects on the edges of
the polyhedrons: cracks of both monolayers
 at the spontaneous curvature of monolayer $C_{\text {0}}<0$,
and a crack of the inner monolayer at $C_{\text {0}}\ge0$. 
Around the latter defect,
the inner monolayer curves positively.
Our results suggested that the polyhedral morphology is controlled by $C_{\text {0}}$.

\end{abstract}


\maketitle

\begin{figure}
\includegraphics{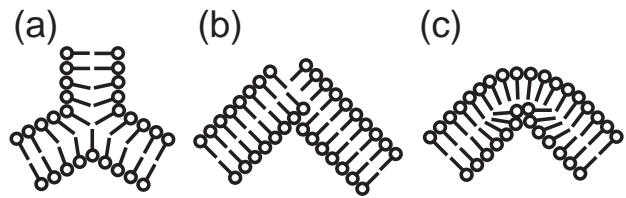}
\caption{ \label{fig:de}
Three types of Line defects. (a) Tilt deformation in the inverted hexagonal phase.
(b) Cracks of both monolayers. Hydrophobic segments are partially exposed. (c)
Crack of inner monolayer. Amphiphilic molecules in the inner monolayer
tilt with respect to the boundary surfaces of two monolayers.
}
\end{figure}

Amphiphilic molecules such as lipids and detergents
form
 various structures such as 
 micelles, cylindrical structures, and bilayer membranes
in aqueous solution~\cite{lip1995}.
In particular,
closed bilayer membranes, i.e., vesicles, 
 are  biologically important as model systems for the
plasma membrane and intracellular compartments in living cells.
Various morphological changes in the vesicles are understood via the 
Helfrich elastic model~\cite{lip1995,helf73,seif97}. 
However, this model can not be applied to non-bilayer structures.
For example, in an inverted hexagonal $H_{II}$ phase,
the hydrophobic interstice (void) space opens
among three cylindrical monolayers.
Recently, it is considered that
this interstice space is filled 
by the tilt deformation of amphiphilic molecules as shown in
Fig.~\ref{fig:de}(a)~\cite{tilts}.
The molecules tilt with respect to the monolayer surfaces around the
junction of the three bilayers.
The monolayer surfaces are sharply bent at the junction.
The effects of tilt deformation are also studied 
regarding the fusion intermediates of fluid phase membranes~\cite{fus}.

On the other hand,
polyhedral-shaped vesicles of $\mu$m scale size were observed
in the gel phase; a triangular pyramid or prism-shaped vesicle of
a mono-component lipid~\cite{sack94}, and an icosahedral vesicle of mixtures of cationic and
anionic surfactants~\cite{dubo01}.
The membranes are flat on the faces of polyhedrons
and are sharply bent at the edges.	
Because the bending modulus are very large
in the gel phase,
the polyhedral vesicles would be stabler than the spherical vesicles.
The free-energy loss of the defects at the edges would be less than
the loss of the equal bending of membranes on the sphere.
However, the defect structure at the edges is unresolved.
The information of the edge structure is significant
to control the morphology of the polyhedral vesicles.
These vesicles are expected to be of practical value for drug delivery.

To clarify the edge structure,
theories or simulations with molecular resolution are needed.
Since molecular dynamics simulations with atomic resolution
have been applied only for the $\sim10$ns 
dynamics of $1000$ lipid molecules 
due to the restrictions of computational time~\cite{mds},
coarse-grained molecular
simulations~\cite{cmds,ber1996,nog2001a, nogs,yama02} have
been applied.
We studied the fusion and fission dynamics of vesicles using Brownian
dynamics~\cite{nogs}.
The self-assembly
into vesicles is simulated by our model~\cite{nog2001a}, a lattice Monte Carlo method~\cite{ber1996}, and dissipative particle
dynamics~\cite{yama02}.
However, these simulated vesicles were flexible, and
no polyhedral vesicle has been obtained.

In the present paper,
we developed our previous model to control the bending modulus of monolayers
by the addition of the curvature potential of a monolayer.
We simulated the polyhedral vesicles of rigid membranes, and
obtained two types of defects at the edges as shown in Figs.~\ref{fig:de}(b) and
(c).
The morphology of the polyhedral vesicles and the defect type depend
on the spontaneous curvature of the monolayer $C_{\text 0}$.

\begin{figure*}
\includegraphics{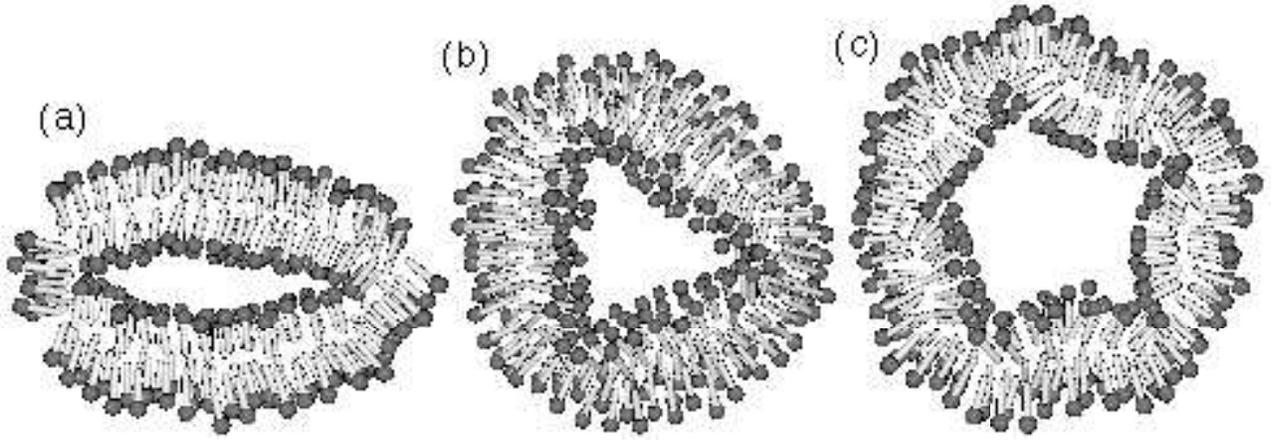}
\caption{ \label{fig:sn}
Sliced snapshots of vesicles at temperature $k_{\text B}T/\varepsilon=0.2$ and the number of
molecules $N=1000$.
(a) Disk shaped (dihedral) vesicle at the spontaneous curvature of monolayer $C_{\text 0}\sigma=-0.11$.
(b) Triangular-pyramid shaped (tetrahedral) vesicle at $C_{\text 0}\sigma=0.058$.
(c) Pentagonal-prism shaped (septihedral) vesicle at $C_{\text
0}\sigma=0.23$.
 Gray spheres and white cylinders
 represent hydrophilic and hydrophobic segments of amphiphilic
 molecules, respectively. 
}
\end{figure*}

An amphiphilic molecule is modeled as rigid rods consisting of one hydrophilic segment ($j=1$)
 and two hydrophobic segments ($j=2, 3$), which are separated by a fixed
 distance $\sigma$. 
Solvent molecules are not explicitly taken into account, and ``hydrophobic'' interaction is mimicked by the multibody local density potential of the hydrophobic segments. 
As details of the basic model
were described in our previous papers~\cite{nog2001a,nogs},
here we briefly explain the model.
The motion of the $j$th segment of the $i$th molecule
follows the underdamped Langevin equation.
Amphiphilic molecules ($i=1,..,N$) interact
 via a repulsive soft-core potential $U_{\text {rep}}$, an attractive ``hydrophobic''
 potential $U_{\text {hp}}$, and a curvature
 potential $U_{\text {CV}}$.
\begin{eqnarray}
U= \sum_{i \ne i'} U_{\text {rep}}(|{\bf r}_{i,j} - {\bf r}_{i',j'}|) 
+  \sum_{j=2,3} U_{\text {hp}}(\rho_{i,j}) 
+ U_{\text {CV}},
\end{eqnarray} 
where $\rho_{i,j}$ is the 
local density of the hydrophobic segments for the $j$th segment of the $i$th molecule. 
Both segments have the same soft radius,
$U_{\text {rep}}(r)/ \varepsilon = \exp \{-20(r/\sigma - 1)\}$.
The ``hydrophobic'' interaction is
 mimicked by the potential $U_{\text {hp}}(\rho)$.
\begin{eqnarray}
\rho_{i,j}= \sum_{i \ne i', j'=2,3} h(|{\bf r}_{i,j} - {\bf r}_{i',j'}|),
\end{eqnarray} 
\begin{eqnarray*}
 {\text{where}} \hspace{0.5cm} h(r)= \frac{1}{\exp \{20(r/\sigma-1.9)\} +1}.
\end{eqnarray*} 
$\rho_{i,j}$ is the number of
 hydrophobic segments in the sphere whose radius is approximately
 $1.9\sigma$.
\begin{equation}
U_{\text{hp}}(\rho)/\varepsilon = \left\{
\begin{array}{ll}
-0.5\rho & (\rho < \rho^*-1) \\
0.25 (\rho-\rho^*)^2 -c & ( \rho^*-1 \leq \rho < \rho^*) \\
-c & (\rho^* \leq \rho) 
\end{array}     
  \right. ,
\end{equation} 
where $c$ is given by $c=0.5\rho^*-0.25$.
We used the values $\rho^* =10$ and $14$ at $j=2$ and $3$, respectively.

To give
the bending modulus $\kappa$ and the spontaneous curvature $C_{\text
{0}}$ of the monolayer membranes,
we use the potential $U_{\text {CV}}$ of the
 orientational difference of neighboring molecules.
\begin{equation}
U_{\text {CV}}= \sum_{i \ne i'}
0.5\kappa'_{\text {cv}} h(r_{i,i'}) ({\bf u}_{i}-{\bf u}_{i'} - C'_{\text {0}} 
 {\bf \hat{r}}_{i,i'})^2,
\end{equation} 
where the vector ${\bf u}_{i}$ is the unit
 orientational vector of the $i$th molecule, and
 ${\bf \hat{r}}_{i,i'}$ ($r_{i,i'}$) is the unit
vector (distance) between the $i$th and $i'$th molecules:
${\bf u}_i = ({\bf r}_{i,1}- {\bf r}_{i,3})/|{\bf r}_{i,1}- {\bf
  r}_{i,3}|$ and ${\bf \hat{r}}_{i,i'} = ({\bf r}_{i}- {\bf r}_{i'})/|{\bf r}_{i}- {\bf
  r}_{i'}|$, where ${\bf r}_{i}$ is the center of mass of the $i$th
molecule.
At $C'_{\text {0}}=0$, this potential is similar to the bending
 elastic potential used in the tethered
 membrane models~\cite{teths}.
When the orientational vectors ${\bf u}_{i}$ are equal to the normal
 vectors of the monolayers
 with no tilt deformation, 
\begin{equation}
U_{\text {CV}}/\varepsilon= \int  0.5\kappa_{\text {cv}}[
(C_{\text {1}}+C_{\text {2}}-C_{\text {0}})^2 - 2 C_{\text {1}}
C_{\text {2}} + C_{\text {0}}^2] dA
\end{equation} 
in the continuum limit, where
$C_{\text {1}}$ and $C_{\text {2}}$ are the two principal curvatures of
a monolayer.
The spontaneous curvature $C_{\text
{0}}$ equals to $C'_{\text {0}}\sigma/\bar{r}_{\text {nb}}$,
where $\bar{r}_{\text {nb}}$ is the mean distance between neighboring
 molecules and $\bar{r}_{\text {nb}}=1.5\sigma$.
On the assumption of the hexagonal packing of molecules in the monolayers,
we obtain $\kappa_{\text {cv}}= \sqrt{3} \kappa'_{\text {cv}}$.
We used $\kappa'_{\text {cv}}=3\varepsilon$ to represent the rigid membrane, and
$\kappa_{\text {cv}}\simeq5\varepsilon$. 
In previous papers, we estimated the bending modulus
$\kappa_{\text 0}/\varepsilon\simeq0.5$ ( half of the bending modulus of bilayers)
at $\kappa'_{\text {cv}}=0$~\cite{nogs}.
Since $\kappa_{\text {cv}}$ is ten times larger than $\kappa_{\text 0}$,
the bending elasticity is mainly given by $U_{\text {CV}}$, and
the bending modulus of the monolayer $\kappa\simeq \kappa_{\text {cv}}$.

We mainly used the number of molecules
$N=1000$ and $k_{\text B}T/\varepsilon=0.2$,
where $k_{\text B}$ is the Boltzmann constant and $T$ is the
temperature.
Amphiphilic molecules spontaneously form vesicles in a fluid phase at $k_{\text
B}T/\varepsilon=0.2$ and $\kappa'_{\text {cv}}=0$.
The unit length $\sigma$ corresponds to $\sim 1$nm. The unit time
step $\tau_{\text 0}=\zeta \sigma^2/\varepsilon$ corresponds to $\sim 1$ns 
estimated from the lateral diffusion constant of
phospholipid at $30^{\circ}$C, $\sim10^{-7}$cm$^2$/s
~\cite{difs},
where $\zeta$ is the friction constant of the segments of molecules.

\begin{figure}
\includegraphics{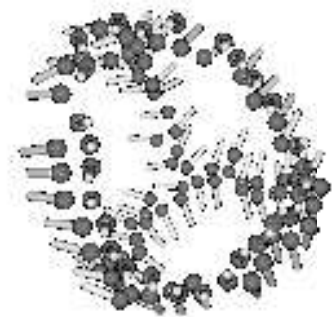}
\caption{ \label{fig:bo}
Line defects (edges) of the tetrahedral vesicle in Fig.~\ref{fig:sn}(b).
The snapshot is viewed from the same viewpoint.
The molecules with number of neighboring molecules $n_i^{\text
{nb}}<4.5$ in the inner monolayer are shown.
The number $n_i^{\text {nb}}$ is defined 
as $n_i^{\text {nb}}=\sum h(r_{i,i'})$.
}
\end{figure}

Vesicles exhibit various polyhedral morphologies at $\kappa'_{\text {cv}}=3\varepsilon$.
The number of faces $n_{\text {f}}$ of polyhedron increases as $C_{\text {0}}$ increases.
Figure \ref{fig:sn} shows examples of the polyhedral vesicles.
The edges of the polyhedrons are formed by the line defects
[Figs.~\ref{fig:de}(b) and (c)].
The molecules at
the line defects are shown using the number of neighboring molecules $n_i^{\text
{nb}}$ in Fig.~\ref{fig:bo}.
The inner monolayers are divided into $n_{\text {f}}$ faces by the
defects.
At $C_{\text {0}}\ge0$, the cracks of the inner monolayer [Fig.~\ref{fig:de}(c)]
occur on the edges, and the outer monolayer consists of one curved
face [Figs.~\ref{fig:sn}(b) and (c)].
At $C_{\text {0}}<0$,  the outer monolayer also exhibits cracks [Fig.~\ref{fig:de}(b)]
on the edges.
When the deformation of
 the left or right side of Fig.~\ref{fig:sn}(a) is formed on the entire
circular-line defects,
the outer monolayer of the disk-shaped vesicles is divided
into three faces (two disks and one cylinder) or
two faces, respectively.

\begin{figure}
\includegraphics{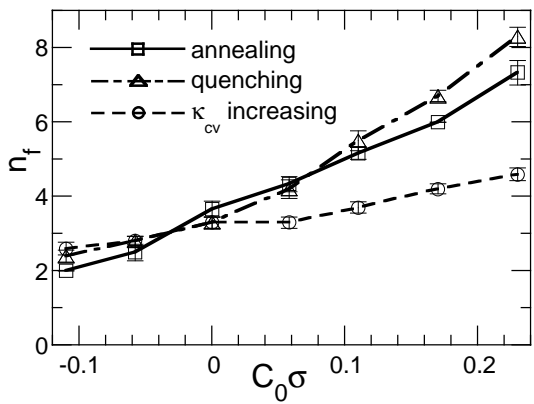}
\caption{ \label{fig:nf}
Spontaneous curvature $C_{\text 0}$ 
dependence of the mean number of faces $<n_{\text f}>$ of polyhedrons
 at $N=1000$ and $k_{\text B}T/\varepsilon=0.2$.
Annealing: Vesicles are annealed from $k_{\text B}T/\varepsilon=0.5$ to $0.2$.
Quenching: Vesicles are simulated starting with spherical
vesicles at $\kappa'_{\text {cv}}=0$.
$\kappa_{\text {cv}}$ increasing: the coefficient $\kappa'_{\text {cv}}$
slowly increases from $\kappa'_{\text {cv}}/\varepsilon=0$ to $3$
starting with the spherical
vesicles at $\kappa'_{\text {cv}}=0$.
}
\end{figure}

Figure \ref{fig:nf} shows the mean number of faces $<n_{\text {f}}>$ at
$k_{\text B}T/\varepsilon=0.2$ using
three methods.
The results of annealing are the closest to the equilibrium values.
Through the other methods,
however, vesicles are often trapped in metastable states.
At $C_{\text {0}}>0$,
vesicles with larger (smaller) $n_{\text {f}}$ values are obtained through quenching
($\kappa_{\text {cv}}$ increasing) than through annealing.
Thus, one should use the annealing method 
to obtain regular polyhedrons.

Flip-flop motion, which is the transverse motion between the inner and
outer monolayers, 
frequently occurs at $k_{\text B}T/\varepsilon=0.5$.
The number of molecules in the inner monolayer
decreases with an increase in $n_{\text {f}}$ at $k_{\text B}T/\varepsilon=0.5$, since
tilting molecules on the line defects share a larger area.
The ratios $\gamma_{\text {in}}$ of molecules in the inner monolayer are $0.31(\pm0.01)$
 and $0.292(\pm 0.003)$ at $C_{\text
{0}}\sigma=-0.11$ ($n_{\text {f}}=2$)
 and  $C_{\text
{0}}\sigma=0.23$ ($n_{\text {f}}=7.3$), respectively.
On the other hand, flip-flop motion
rarely  occurs, and the ratio $\gamma_{\text {in}}$ is fixed
at $\gamma_{\text {in}}=0.328(\pm0.003)$ at the
fixed temperature $k_{\text B}T/\varepsilon=0.2$.
In typical experimental conditions,
flip-flop motion is very slow, and the half-life is more than several
 hours even in the fluid phase~\cite{flis}.
Thus experimentally,
the ratio $\gamma_{\text {in}}$ of the polyhedral vesicles
should not reach an equilibrium value as well as the simulation with the
fixed temperature $k_{\text B}T/\varepsilon=0.2$.

\begin{figure}
\includegraphics{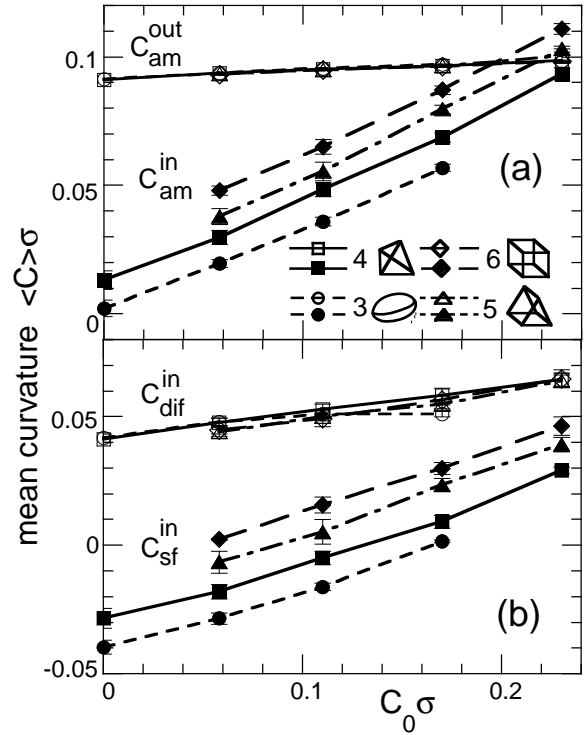}
\caption{ \label{fig:sc}
Spontaneous curvature $C_{\text 0}$ 
dependence of (a) the mean curvature for amphiphilic molecules,
$<C_{\text {am}}>$
(b) the mean curvature for monolayer surface $<C_{\text {sf}}>$
and the difference $<C_{\text {dif}}>=<C_{\text {am}}-C_{\text {sf}}>$
 at $N=1000$, $k_{\text B}T/\varepsilon=0.2$, and $\gamma_{\text {in}}=0.328(\pm0.003)$.
The superscripts 'in' and 'out' represent inner and outer monolayers,
respectively. 
Circles: rugby-ball shaped trihedron. Squares: tetrahedron.
Triangles: triangular prism.
Diamonds: cube.
}
\end{figure}

In the obtained polyhedrons,
three edges are connected at any vertex.
The connections of more edges are unstable
 and are not formed.
The number of edges $n_{\text e}$ then equals to $1.5$ times the
number of vertices $n_{\text v}$, 
because each edge contacts two vertices.
We derived $n_{\text e}= 3(n_{\text f}-2)$ and $n_{\text v}= 2(n_{\text f}-2)$
from this relation and
Euler's formula for a convex polyhedron ($n_{\text f}+ n_{\text
v}-n_{\text e}=2$).
At $n_{\text f}\ge6$, multiple types of polyhedrons with the same
number of faces exist.
However, we obtained only one or two types of polyhedrons; i.e., 
cube ($4^6$) at $n_{\text f}=6$; pentagonal prism ($4^5,5^2$) and ($3^1,4^3,5^3$)
 at $n_{\text f}=7$; ($4^4,5^4$) at $n_{\text f}=8$; ($4^3,5^6$)
and ($4^4,5^4,6^1$)
at $n_{\text f}=9$, where ($p^q$) represents a polyhedron with $q$ $p$-gons.
Thus more symmetric polyhedrons are more frequently formed.

Figure \ref{fig:sc}(a) shows the $C_{\text {0}}$ dependence of 
the curvatures of four polyhedrons at $C_{\text {0}}\ge0$,
where
$C_{\text {am}}$ is the curvature (splay) for amphiphilic molecules,
$<C_{\text {am}}>\bar{r}_{\text {nb}}= (\sum ({\bf u}_{i}-{\bf u}_{i'}) 
 {\bf \hat{r}}_{i,i'} h(r_{i,i'}))/(\sum h(r_{i,i'}))$.
The mean curvature $<C_{\text {am}}^{\text {out}}>$ of the outer monolayer
is almost independent of $C_{\text {0}}$ and $n_{\text {f}}$.
On the other hand, $<C_{\text {am}}^{\text {in}}>$ of the inner monolayer
increases with an increase in $C_{\text {0}}$ and $n_{\text {f}}$,
since the inner monolayers curve positively 
around the crack of the inner monolayers.
The mean curvature $<C_{\text {am}}^{\text {in}}>$ for a larger $n_{\text {f}}$ is closer to $C_{\text {0}}$,
though more hydrophilic segments contact
the hydrophobic segments at the line defects.
The curvature $C_{\text {am}}$ does not coincide with the curvature of
the monolayers, since molecules can tilt with respect to the monolayer surface.
To clarify this tilt, 
we estimated the curvature $C_{\text {sf}}$ for the monolayer surface,
$<C_{\text {sf}}>\bar{r}_{\text {nb}}= (\sum ({\bf n}_{i}-{\bf n}_{i'}) 
 {\bf \hat{r}}_{i,i'} h(r_{i,i'}))/(\sum h(r_{i,i'}))$,
where ${\bf n}_{i}$ is the normal vector of the monolayer surface at
${\bf r}_{i}$.
We defined ${\bf n}_{i}$ as the vector minimizing $\varepsilon_i= \sum
h(r_{i,i'}) ({\bf n}_{i} {\bf \hat{r}}_{i,i'})^2$ when $n_i^{\text {nb}}>2.5$.
This minimizing vector is the eigenvector
with the smallest eigenvalue of the moment tensor of inertia of the
neighboring molecules. 
Figure \ref{fig:sc}(b) shows 
the curvatures $<C_{\text {sf}}^{\text {in}}>$
of the inner monolayer surface and difference $<C_{\text {dif}}^{\text
{in}}>$ between the two curvatures.
The inner monolayer surface (molecules in the inner monolayer) 
tilt with respect to the boundary surfaces
of the two monolayers (the inner monolayer).
Both tilts increase $C_{\text {am}}$.
Since the molecular tilt in the inner monolayer $<C_{\text {dif}}^{\text
{in}}>$ is almost independent of $n_{\text {f}}$,
the length of the line defects only changes the curvature $<C_{\text {sf}}^{\text {in}}>$
of the inner monolayer surface.
Thus the polyhedral morphology at equilibrium should be determined
by the effects of line defects on $<C_{\text {sf}}^{\text
{in}}>$ and the hydrophobic interaction.

We found that the spontaneous curvature $C_{\text 0}$ can be estimated using
this tilt deformation with respect to the monolayer surface.
At $C_{\text {sf}}<C_{\text 0}$ ($C_{\text {sf}}>C_{\text 0}$),
molecules tilt to reduce $|C_{\text {am}}-C_{\text 0}|$ 
and $C_{\text {dif}}$ shows a positive (negative) value.
Thus, we estimate $C_{\text 0}\sigma\simeq0.04$ at
$\kappa'_{cv}=0$ and $k_{\text B}T/\varepsilon=0.2$:
$<C_{\text {dif}}>\sigma=0.0041(\pm 0.0002)$ for flat membranes;
$<C_{\text {dif}}^{\text {out}}>\sigma=0.0(\pm 0.00007)$ and $<C_{\text {sf}}^{\text {out}}>\sigma=0.0429(\pm 0.0005)$ for tube-shaped vesicles with a diameter of
$30\sigma$; and $<C_{\text {dif}}^{\text {out}}>\sigma=-0.0056(\pm 0.0003)$ and
 $<C_{\text {sf}}^{\text {out}}>\sigma=0.0975(\pm 0.0003)$ for
spherical vesicles with a diameter of $20\sigma$.

The line defects may be interpreted using the 
correction terms of the Helfrich model,
the local minimum at a large $C_{\text 1}+C_{\text 2}$.
The morphology of the polyhedral vesicles may then be obtained from the Euler-Lagrange
differential equation.
Similar deformations to the crack of the inner monolayer are seen in our daily experience.
When a rubber hose is strongly bent,
a side of the hose becomes hollow, and the other side
 smoothly bends.

The morphology of the polyhedral vesicles
depends on the number $N$ and properties of molecules.
At $N=2000$ and $C_{\text {0}}\sigma=0.23$,
vesicles exhibit more complex morphologies with
concave edges, where the membrane bends outside
with a crack of the outer monolayer.
The hexagonal packing of molecules should stabilize
the triangular and hexagonal faces.
In some multi-component vesicles,
the phase separation occurs at the edges or vertices of the polyhedrons.
Dubois {\it et al.} reported that the segregated anionic surfactants
form pores at the vertices~\cite{dubo01}. 
Various polyhedral vesicles are likely to be
experimentally observed under
the control of $C_{\text {0}}$ and other conditions.

This work was supported in part by a Grant-in-Aid for Scientific Research from the Ministry of Education, Culture, Sports, Science, and Technology of Japan.


\begin{references}
\bibitem{lip1995} 
{\it Structure and Dynamics of Membranes} edited by R. Lipowsky and E. Sackmann
(Elsevier Science, Amsterdam, 1995); S. A. Safran, {\it Statistical
Thermodynamics of Surfaces, Interfaces, and Membranes} (Addion-Wesley,
New York, 1994).
\bibitem{helf73} W. Helfrich, Z. Naturforsch {\bf 28c}, 693 (1973).
\bibitem{seif97} U. Seifert, Adv.\ Phys. {\bf 46}, 13 (1997).
\bibitem{tilts} M. Hamm and M. M. Kozlov, Eur. Phys. J. B {\bf 6}, 519 (1998).
\bibitem{fus} P. I. Kuzmin, J. Zimmerberg, Yu. A. Chizmadzhev,
 and F. S. Cohen,  Proc.\ Natl.\ Acad.\ Sci.\ USA {\bf 98}, 7235
(2001); Y. Kozlovsky and M. M. Kozlov, Biophys.\ J. {\bf 82},
882 (2002).
\bibitem{sack94} E. Sackmann, FEBS lett. {\bf 346}, 3 (1994).
\bibitem{dubo01} M. Dubois {\it et al.}, Nature {\bf 411}, 672 (2001).
\bibitem{mds} L. R. Forrest and M. S. P. Sansom, Curr.\ Opin.\
 Struct.\ Biol. {\bf 10}, 174 (2000);
S. E. Feller, Curr.\ Opin.\ Colloid\ Interface\ Sci.
 {\bf 5}, 217 (2000);
S. J. Marrink, E. Lindahl, O. Edholm, and
 A. E. Mark, J.\ Am.\ Chem.\ Soc. {\bf 123}, 8638 (2001).
\bibitem{cmds} S. Karaborni {\it et al.}, Science {\bf 266,} 254
(1994); R. Goetz, G. Gompper, and R. Lipowsky,
 Phys.\ Rev.\ Lett. {\bf 82}, 221 (1999);
T. Soddemann, B. D\"unweg, and K. Kremer, Eur. Phys. J. E {\bf 6},
409 (2001); R. D. Groot and K. L. Rabone, Biophys.\ J. 
{\bf 81}, 725 (2001).
\bibitem{nog2001a} H. Noguchi and M. Takasu, Phys.\ Rev.\ E {\bf 64},
 041913 (2001).
\bibitem{nogs} H. Noguchi and M. Takasu, J.\ Chem.\ Phys. {\bf 115},
 9547 (2001); Biophys.\ J. {\bf 83}, 299 (2002); Phys.\ Rev.\ E {\bf
65}, 051907 (2002); H. Noguchi, J.\ Chem.\ Phys. (2002) in press.
\bibitem{ber1996} A. T. Bernardes, J.\ Phys.\ II {\bf 6}, 169
 (1996).
\bibitem{yama02} S. Yamamoto, Y. Maruyama, and S. -a. Hyodo,
J. Chem. Phys. {\bf 116}, 5842 (2002).
\bibitem{teths} H. S. Seung and D. R. Nelson, Phys. Rev. A {\bf 38},
1005 (1988); G. Gompper and D. M. Kroll, Phys. Rev. E {\bf 51}, 514 (1995).
\bibitem{difs} E.-S. Wu, K. Jacobson, and D. Papahadjopoulos, 
Biochemistry {\bf 16}, 3936 (1977);  W. Pfeiffer {\it et al.},
Europhys. Lett. {\bf 8}, 201 (1989).
\bibitem{flis} 
R. D. Kornberg and H. M. McConnell, Biochemistry {\bf 10}, 1111 (1971);
P. F. Devaux, Biochemistry {\bf 30}, 1163 (1991).
\end{references}
\end{document}